# A Data- and Task- Oriented Design Framework for Bivariate Communication of Uncertainty

Letitia Sabburg, Alan Woodley and Kerrie Mengersen

**Abstract**—The communication of uncertainty of estimates, predictions and insights based on spatio-temporal models is important for decision-making as it impacts the utilisation and interpretation of information. Bivariate mapping is commonly used for communication of estimates and associated uncertainty; however, it is known that different visual qualities resulting from choices of symbols and consequent interaction between the display dimensions can lead to different interpretations and consequently affect resultant decisions. Characteristics of the data to be presented, such as spatial format, statistical level, and continuousness, shape the range of available bivariate symbols. The subsequent utility of these bivariate symbols depends on their ability to achieve the end-user's goals. In this paper we present a novel design framework, which, through consideration of both input data characteristics and potential operational tasks (as proxy to end-user goals), assists map designers in appropriate selection of bivariate symbols for the coincident presentation of spatio-temporal modelled data and associated uncertainty. The framework is showcased through application to a case study pertaining to sediment pollution in the Great Barrier Reef.

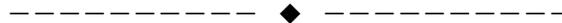

## 1 INTRODUCTION

In the context of decision-making, spatio-temporal modelled estimates and predictions, and associated information about the uncertainty of these values, are important to communicate collectively. This is because when dealing with imperfect information, such as model outputs, associated uncertainty estimates can greatly influence conclusions drawn from model predictions and highlight where further investigation is required, enhancing the decision-making process [19]. If uncertainty estimates associated with spatio-temporal model predictions are not provided, decision-makers cannot assess the reliability of predictions and effectiveness of decisions based on model outputs [18], [24]. Therefore, when properly communicated, uncertainty can help make better informed decisions [34].

Visualisations are often used to condense individual pieces of information so that complex ideas can be comprehended more easily [19]. This has led to increased interest in incorporating both point estimates and uncertainty estimates together in maps [3], [4]. However, the choice of visualisation method directly impacts our ability to comprehend complex and simultaneous information, particularly in space-time [11]. Suboptimal presentation, whether through inappropriate layout or misrepresentation of the data, can lead to lengthy and/or incorrect interpretation, consequently impacting any resulting decisions [12], [29], [34].

Due to the wide range of abilities, preferences and requirements of the users, and the diverse characteristics of data and outputs to be presented, there is a lack of consensus on best methods for visualising predictions and estimates of uncertainty together [31], [34]. Consequently a new kind of semiotic framework in bivariate uncertainty visualisation is needed [15]. This requires finding an interface or mapping between the data to be visualised, the end-users goals and potential uncertainty representations [17], [25]. As such, the design of an appropriate bivariate uncertainty representation is a data- and task- oriented problem.

This paper presents such a data- and task- oriented bivariate design framework. To showcase this framework, the remainder of this paper is split into sections containing an initial investigation into related works, a description of the proposed design framework, a demonstration of the framework as applied to a case study, and subsequent discussion on findings and limitations both in context of the case study and more generally.

## 2 RELATED WORKS

### 2.1 Communicating Uncertainty
Kinkeldey et al. [15] characterise the possible methods of visual signification of uncertainty into a set of five dichotomies, outlined below:

1. intrinsic or extrinsic display of uncertainty,

Intrinsic display methods are where graphic variables are physically aligned with data's spatial boundaries. Alternatively, information can be visualized extrinsically through use of glyphs such as error bars, pie charts, or

────────────────
- *Letitia Sabburg is with Queensland University of Technology, e-mail: l.sabburg@hdr.qut.edu.au.*
- *Alan Woodley is with Queensland University of Technology, e-mail: a.woodley@qut.edu.au.*
- *Kerrie Mengersen is with Queensland University of Technology, e-mail: k.mengersen@qut.edu.au.*

other multivariate point symbols [12].

2. coincident or adjacent mapping,

Adjacent mapping is where information is presented through a juxtaposition, on separate side-by-side visualisations. Coincident mapping is where a second visual channel is introduced to the display to include another set of information [5].

3. static or dynamic presentation,

Static visualisations are those which remain constant. Presentation methods can also be interactive or dynamic such as through animations of data or map ensembles, virtual reality (VR) technology or web mapping, allowing for user-controlled display of information [4], [9], [40].

4. integral or separable interpretation of uncertainty,

When spatio-temporal data can be interpreted independently of its uncertainty then the two sets of information are separable; however, if the uncertainty cannot be ignored then the information sources are integral to one another [26].

5. implicit or explicit communication of uncertainty.

Explicit communication conveys uncertainty as a direct measurement/value. Implicit communication is where uncertainty is inferred from the range of possible results [22].

Intrinsic methods are typically preferred by decision-makers [22] and coincident mapping allows for options of both integral and separable communication of variables, which as discussed in Section 2.2, is an important consideration in task-oriented bivariate semiotic design. There is however little evidence in recommendation of dynamic representation over static as there are so many different types of dynamic approaches, making it hard to come to consistent conclusions about their effectiveness [15]. There is also limited support either way for explicit or implicit communication of uncertainty as these methods result in different decisions [7], therefore their adoption depends on user-task requirements. However, explicit methods are commonly adopted and, similarly to coincident mapping, allow for both integral and separable interpretation of variables. As such, the remainder of this research focuses on the intrinsic, coincident, static, and explicit signification of uncertainty through appropriate design of bivariate symbology.

## 2.2 Aspects of Designing a Bivariate Symbol

*Visual Variable Perception*

Visual variables or cues describe a fundamental set of visual options a designer can choose to effectively display information in graphics [37]. This includes symbol properties such as size, hue, value, shape, texture, orientation, saturation, arrangement, transparency and blur [2], [11], [20], [25], [37]. Visual variables can be complex having constituent dimensions [25]. For example, pattern characteristics are defined through a combination of shape, orientation, arrangement, spacing, size, and texture [25], [30], [32], [39]. As human colour perception lacks orthogonality (that is, the properties of colour affect perception of one another), colour channels (hue, saturation and value) must also be viewed holistically [5], [11], [37]. Cues can also be functionally related, for example ink coverage changes with respect to a change in pattern size or density [10].

Bertin [2] presented a set of perception properties which describe the behaviour of visual variables, outlined below:

1. associative perception is where the visibility of each sign is the same;
2. selective perception is where there is an ability to immediately isolate all elements of a given category;
3. ordered perception means there is an immediate/intuitive recognisability of ordering within visual cue;
4. quantitative perception allows perceptibility of numeric ratios; and
5. length refers to the number of perceptually distinguishable steps of the selective perception [2].

A cues' length is not necessarily autonomous. For example the length of value varies by hue; a yellow hue only facilitates distinguishing three levels of value, however this increases to 6 or 7 for red and blue hues [17]. Denser patterns also have less perceptibility between levels, and symbols become less distinct with more variety [11], [17]. Conversely, as the number of cue levels used decreases, the perceptibility between levels and consequent strength of selectivity increases [17].

Visual variables also have a perceptual ordering, with dominant cues being more effective for visual processing [23]. Bertin [2] explains that size and value tend to dominate most other cues because of their dissociative nature; consequently, value dominates the interpretation of colour.

Visual elements on a map can also interfere with one another, changing our interpretation of the presented information [17]. In bivariate maps, this is particularly noticeable as interactions between display dimensions. A pattern's texture for example can appear distorted when overlaid on a different texture, making it appear finer or coarser than the illustrator intended [37]. Additionally, as the interpretation of colour is relative, colours can be perceived differently depending on their surrounds, or they may visually blend such as in the case of fine textured pattern overlays and their background colour [2], [11].

Perception of visual variables also impacts users' abilities to focus on specific aspects of the data. Roth [32] investigated how a range of symbol properties catch the reader's eye; this is typically seen within dissociative cues. It should also be noted that our attention is generally drawn to what is different from the norm, therefore, what stands out depends on the data's distribution and resultant map composition [19].

*Separability of Visual Variable Combinations*

Combinations of visual variables can be used for redundant symbolization, to enhance graphic encoding of an individual attribute, or in bivariate communication [32].
In bivariate representations, interactions between visual dimensions can influence the interpretation of information; as our interpretation of one visual variable typically comes at the expense of another, there are usually trade-offs when using bivariate symbols [5], [26], [37]. Consequently, understanding how these visual dimensions behave together is crucial for effective bivariate map symbolisation.

Bivariate symbol interactions facilitate either divided or selective attention characterised by their standing on the integral-separable continuum [26], [32]. Separable dimensions do not interfere with one another's interpretation and are therefore, appropriate for displaying spatially coincident but unrelated data, that allow independent retrieval of each variable's characteristics [26], [27], [28], [37]; in this case, the scales of each attribute can be incongruous [32]. However, as visual dimensionality increases it becomes difficult to find cue combinations that are perceptually accurate and separable [5], [26].

Integral combinations, due to the inter-dimensional interference, prevent map readers from separating out information. This is useful when the relationship between variables is more important than the individual attributes, seeing how variables co-vary with one another and highlighting any correlations [26], [32], [37].

Pairs of visual cues with aspects of both integrality and separability are considered configural. These combinations display a new emergent feature that only exists when both values of the bivariate symbol are read together and can be used as a third stronger visual variable [28]. It is recommended to use configural combinations for highlighting interactions whilst retaining variable's individual characteristics, but both variables should have congruous scales [26], [32].

Asymmetric conjunctions occur when one visual dimension dominates over another. As noted above, this typically occurs when one of the visual dimensions is dissociative, such as size or value [10]. Asymmetry creates an emergent visual dimension, which is stronger than the non-dominant base visual variable; consequently, these conjunctions retain some degree of separability [10]. Due to the production of an imbalanced visual effect and tendency for readers to interpret one attribute over another, asymmetry can be useful in situations when one variable is more important than another. It can be used to downplay the visual impact of certain data ranges or meaningfully highlight/dehighlight features [10], [27], [32].

Research by Nelson [27], Retchless and Brewer [30] and Elmer [10] provides a sizeable characterisation of bivariate symbols on the integral-separable continuum. However, many of these are author postulations and remain to be empirically evaluated. Consequently, we do see some variation in opinion across the literature.

*Data Characteristics and Visual Variable Applicability*

The types of information conveyed through a map can be classified into either object, space or time components, with the respective phenomena being described as thematic, spatial or existential property change [8].

For the thematic component, data are generalised from the point of view of the cartographer through categorisation into levels of measurement, that is, nominal, ordinal and numeric; numeric data can also have a subclassification as either interval or ratio [8]. Alternatively, thematic information may be classified more simply by whether it is qualitative or quantitative and/or whether the phenomenon is continuous or discrete [2], [8], [25].

In terms of the spatial component, the convention is to classify the objects and phenomena as either discrete or continuous [8], [20]. Discrete objects are usually further categorised as either point, line or area [8].

When designing a good symbol, the spatial and statistical properties discussed above should be considered in concert. If the representation of the data is discrete rather than continuous, additional consideration must also be given to how many data levels are being presented; as different visual variables support different lengths of selectivity and thus levels of perceptibility [2], [17].

The implication of data characteristics on symbol applicability has been explored by various authors. Bertin [2] related the perceptive properties of cues to their spatial implantation. Slocum [35] ranked the applicability of different visual variables to each level of measurement (nominal, ordinal, numeric) [39]. As an extension to these results, Roth [32] investigated additional cues pertinent to uncertainty representation, as well as juxtaposed their associative and selective nature. Morrison [25] made recommendations for adoption of specific cues against each cross classification of spatial implantation (point, line, or area), statical categorisation (qualitative or quantitative), and map colour style (coloured or black and white). In the context of bivariate symbol design, Nelson [27] examined the nature of cue pairs on the integral-separable continuum with relation to their spatial implantation and qualitative/quantitative nature.

*Bivariate Symbol Configurations*

For bivariate symbologies, colour can be used in presenting both or just one of the data dimensions to be communicated. Due to the nature of colour perception, the performance of a symbol with two colour dimensions is based on the component colour channels (hue, saturation, value) and their relationship, as well as the surrounding colours in the map [5]. This also means that in practice, colour schemes should be minimised in length to keep distinguishability high [5]. Most recently, it is suggested to keep these schemes to no more than nine bivariate levels, to allow for their internalisation [19].

In the context of uncertainty mapping, bivariate colour schemes are often comprised of a multi-hued colour palette in one dimension, with a change in either saturation, value, or transparency in the other. A simple example of this is the diverging colour scheme described by MacEachren et al. [22]. In this scheme the thematic variable is represented using contrasting hues, which, with increasing uncertainty, converge to white through a reduction in colour value. However, under this scheme the thematic variable is limited to binary classification, and generally trends are best seen from colour scales which don't converge/diverge [17].

When colour is only used to encode one variable, a range of other bivariate symbols become available which differ in performance and applicability. Conjunctions of one coloured and one non-coloured cue are typically good for retaining separability [27], [30]. Colour and texture combinations are particularly effective for unexperienced users as they can show uncertainty information without limiting pattern detection of the thematic data [21], [29].

Compared to integral colour schemes, these separable pattern-filled methods are easier to read and better at prompting recognition of reliability of an area. Integral colour schemes, on the other hand, can prevent users seeing uncertainty clusters of information at a glance [21], [30].

Bivariate symbols do not need to have any colour-based display dimension. An interesting example of this is cross-hatch patterns, where the density for each hatching level is used as independent representation of each variable; this method is integral and effective at communicating correlation, but with caveats for small enumerations [30].

## 2.3 Effective Representation of Uncertainty Using Bivariate Symbology

*Semiotic Fit for Uncertainty Communication*

The semiotic fit for representing uncertainty, is commonly rationalised through metrics such as user interpretation accuracy and speed, user preference and confidence, and symbol effectiveness and intuitiveness [5], [22], [30], [34], with each of these factors being controlled by the symbol's perceptive properties and limitations. Cues which are consistently considered intuitive for representing uncertainty across the literature include saturation, value, texture, transparency, blur and resolution [16].

Through two linked empirical studies, MacEachren et al. [23] asserts that fuzziness, location, and value are overall most applicable for representing discrete entity ordinal uncertainty; arrangement, size and transparency are also acceptable, but saturation, hue, orientation, and shape are not. Size and colour value may generally be more appropriate for depicting numerical uncertainty, and texture is preferred for binary representation of uncertainty existence [34].

Kinkeldey et al. [15] summarised other relevant literature, finding that for response accuracy, value outperforms texture, blur and sketchiness, and texture and transparency also outperform saturation; therefore, saturation should only be used if other cues are unavailable. The use of degradation techniques such as blur or resolution can help make certain information stand out, but this comes with the risk of making important information illegible [37].

*Operational Effectiveness of a Visualisation*

The utility of the implemented symbology extends to the user's specific objectives and comprehension abilities [29]. This is because different users may have different goals in mind and their individual perceptions, preferences and expertise all impact their visual interpretation and use of information [16], [34]. Therefore, ultimately the effectiveness of a representation (uncertainty or otherwise) is not just determined by semiotic fit to the data being presented, but also by how it alters interpretation of the data and influences the use of the map for a given purpose [20].

However, this cannot always be immediately determined. Instead, we can infer this measure by how effective the representation is at accomplishing the operational tasks required in achieving end-user goals.

Wehrend and Lewis [38] address a comparable problem by considering the crossed classification of an object being presented, categorised by its statistical and spatial characteristics, with user-operations performed using the output visualisation, assigning candidate visualisation techniques to each 'problem' in the resultant matrix.

Potential operational tasks can be categorised by how the data is used (overview, zoom, filter, detail-on-demand, relations and historical), the cognitive operation required (identify, locate, compare or associate), or with respect to the focus level of the data (described as level of reading by Bertin [2], individual data element, subset of data elements or whole data set) [2], [8].

*Summary of Bivariate Symbols in Representing Uncertainty*

There has been some notable investigation on the applicability of different bivariate symbols for representation of uncertainty coincident with spatio-temporal data. This is summarised below.

Paired with a warm hue colour scheme for the thematic variable, Retchless and Brewer [30] used ranking tasks to test a range of classed and unclassed colour-related cues (such as chroma; chroma and value) and various pattern styles as uncertainty representation in a bivariate scheme. None of the pattern styles differed significantly in performance, but all colour-related cues performed significantly worse, and overall suitability of the different bivariate techniques varied with respect to the map maker's goals. The results also suggested that when using colour-related cues in both display dimensions, interpretation of uncertainty improves when its representation is classed and has a lightness component (value). In contrast, this worsened the interpretation of the thematic variable, likely because lightness was a component across both variables, making the information hard to separate. Retchless and Brewer [30] also noted that by using a conjunctive hue-value cue for the thematic variable and conjunctive saturation-value cue for the uncertainty variable, some amount of separability could be retained.

Lucchesi and Wikle [19] investigated the commonly adopted bivariate colour scheme, where two single hue colour palettes vary in lightness based on their respective data dimension (thematic or uncertainty) and are then mixed to produce a coloured grid. Although this scheme allows map users to detect trends and interpret numerical information for both variables, any more than nine classes would require memorisation of the legend/key and not intuitive/internalised usage of the bivariate scheme. Lucchesi and Wikle [19] also formulated a novel pixelation method, based around the need of limiting interpretability of highly uncertain areas. This method created a textural difference (smoother or coarser appearance) through the consistency or inconsistency of pixel colours within an area; however, examination of the effectiveness of this method has not yet been performed.

Correll et al. [5] paralleled the cool hue series (for the thematic variable) and saturation (for uncertainty) but with a non-uniform budgeting of visual channels; a scheme designed to suppress uncertain values for risk-adverse decision making. This scheme increases discriminability between, and interpretability of, more certain information, due to the increased availability of visual classes to higher certainty data. Their research highlighted how uncertainty presentation has a measurable impact on decision making. More specifically, they found perceptual decoding error was high for continuous presentation compared to categorised information, and especially so because of the integrality of colour channels; thus, they recommend employing discrete schemes with a small number of categories.

## 2.4 Summary of Related Works

It can be concluded that in the decision-making context an intrinsic, coincident, static, and explicit style of bivariate signification of uncertainty is likely the most suitable visual communication approach. Under this approach, the applicability, and limitations of adopted visual variables are related to the characteristics of the data being presented (spatial, statistical, continuousness), how the visual variables interact, and specifically for the uncertainty representation, the visual variables' inherent semiotic fit to the data.

Furthermore, there are clear gaps in knowledge of bivariate symbol design, namely, end-user goals and consequent operational tasks are not accounted for in existing design methods. This corroborates the need for a standardised semiotic design methodology for the bivariate communication of uncertainty.

## 3 FRAMEWORK

This paper presents a design framework which jointly considers the characteristics of the data to be presented and the end user's goals. This addresses the lack of standardisation in bivariate symbology design for the intrinsic, coincident, static, and explicit signification of uncertainty.

The framework is structured as the five-step design methodology as presented in Fig. 1.

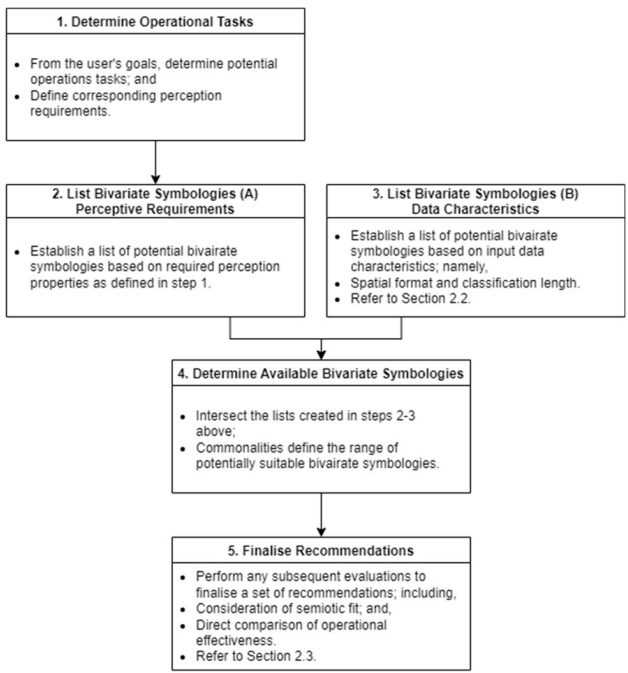

Fig. 1. Five-step methodology for bivariate symbol design.

To utilise the framework in the context of communicating spatio-temporal outputs and associated uncertainty, the design methodology need only consider the ordinal and quantitative representation of information. Moreover, the effects of some visual variables and visual variable combinations on the visual interpretation of data are not well understood. This has limited the consideration of visual variables within the design methodology to those presented in Fig. 2, and consequent bivariate symbologies (organised by spatial implantations) presented in Table 1.

The visual variables presented in Fig. 2. include size, value, saturation, transparency, blur, density, and texture; all of which support ordered perception as is typically required by spatio-temporal estimates. Due to their inherent semiotic intuitiveness in conveying uncertainty, saturation, blur, and transparency are recommended solely for representing the uncertainty dimension. This constraint is carried across into Table 1.

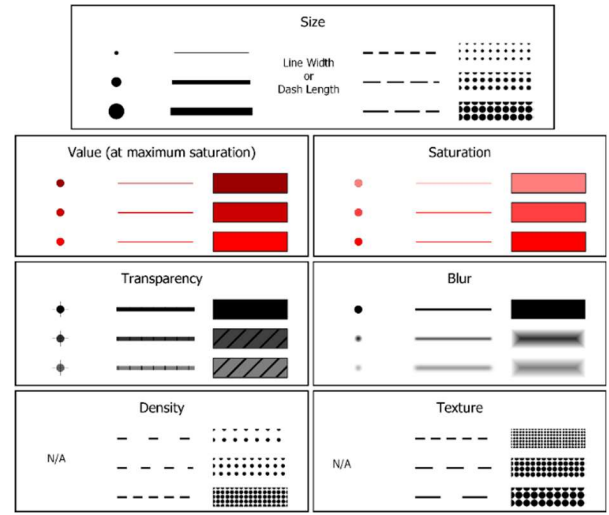

Fig. 2. Graphical representation of employed visual variables.

TABLE 1
BIVARIATE SYMBOLS TYPIFIED BY SPATIAL IMPLANTATION

| (a) Thematic Representation | Point | Uncertainty Representation | | | | | | |
|---|---|---|---|---|---|---|---|---|
| | | Saturation | Blur | Transparency | Value | Size | Texture | Density |
| | Value | Y | Y | Y | Y[A] | Y | | |
| | Size | Y | Y | Y | Y | | | |
| | Texture | | | | | | | |
| | Density | | | | | | | |
| (b) Thematic Representation | Line | Uncertainty Representation | | | | | | |
| | | Saturation | Blur | Transparency | Value | Size | Texture | Density |
| | Value | Y | Y | Y | Y[A] | Y | Y | Y |
| | Size | Y | Y | Y | Y | Y[C] | Y[B] | Y[B] |
| | Texture | Y | Y | Y | Y | Y[B] | | |
| | Density | Y | Y | Y | Y | Y[B] | | |
| (c) Thematic Representation | Area | Uncertainty Representation | | | | | | |
| | | Saturation | Blur | Transparency | Value | Size | Texture | Density |
| | Value | Y | Y | Y | Y[A] | Y | Y | Y |
| | Size | Y | Y | Y | Y | Y[D] | | |
| | Texture | Y | Y | Y | Y | | | |
| | Density | Y | Y | Y | Y | | | Y[D] |

[A]Bivariate choropleth as described by Lucchesi and Wickle [18], [B]Size employed as line width, [C]Size with Size employed as line width with dash length, [D]Each dimesnion applied to a corresponding hatch within a crosshatch symbology.

## 3.1 Contextual Information

The five-step design methodology presented works by simultaneously a) considering the applicability of visual variables in representing data of varying spatial, statistical and continuity traits presented in Section 2.2, and b) addressing the perceptive requirements of employed bivariate symbols in achieving operational tasks, as a proxy to achieving end-user goals.

Tables 2 and 3 support the consideration of data characteristics in bivariate symbology design. Extending from Bertin [2] and Roth [32], Table 2 presents the perception properties, and selective lengths with respect to the spatial implantation, of the visual variables employed within the framework. All considered visual variables have selective and ordered perception, however texture and size are the only variables with associative or quantitative perception, respectively. Additionally, within bivariate application, the selective lengths of either visual dimension is limited to between 3 and 5 levels depending on spatial implantation.

Table 3 presents the classification of employed bivariate symbols on the integral-separable continuum as a synthesis of Nelson [27], Retchless and Brewer [30] and Elmer [10], with additional author inferences. Few characterisations change across spatial implantation. Bivariate schemes where variation in colour occurs across both visual dimensions are always integral. Most other bivariate combinations presented are separable, with limited asymmetric or configural options.

Addressing the perceptive requirements of a given visualisation is achievable using operational task examples provided within Table 4. Table 4 is an adaptation of Wehrend and Lewis [38] and Dykes et al. [8] sets of cognitive operations, Roth and Mattis [33] list of goals and Bertin [2] list of reading levels. The list provides a set of examples to act as a guideline when relating user-tasks, as a proxy to user-goals, to potential symbol perception requirements.

When addressing complex user-goals, the need to support multiple operational tasks is expected. This can potentially result in more complex and/or potentially conflicting symbol perception requirements. Under this circumstance

TABLE 2
PERCEPTIVE PROPERTIES AND SELECTIVE LENGTH UNDER SPATIAL IMPLANTATION OF EMPLOYED VISUAL VARIABLE

|  | Perceptive Property Observance* | | | | Selective Length by Spatial Implantation | | |
|---|---|---|---|---|---|---|---|
|  | Selective | Associative | Ordered | Quantitative | Point | Line | Area |
| Blur | Y | - | Y | - | 3 | 3 | 3 |
| Transparency | Y | - | Y | - | 3 | 4 | 5 |
| Saturation | Y | - | Y | - | 3 | 4 | 5 |
| Value | **Y** | **-** | **Y** | **-** | **3** | **4** | **5** |
| Size | **Y** | - | **Y** | **Y** | **4** | 4 | **5** |
| Texture | **Y** | **Y** | **Y** | **-** | ■ | **4** | **5** |
| Density | Y | - | Y | - | ■ | 4 | 5 |

*Equivalent across all spatial implantations. Values underlined and in bold have been established within literature; all others values are estimates by this author. It is postulated that a) transparency, value, and saturation have equivalent perception properties, b) area density and area size have equivalent selective lengths, and c) blur has no more than three levels of selectivity under any spatial implantation.

TABLE 3
CLASSIFICATION ON THE INTEGRAL-SEPARABLE CONTINUUM OF EMPLOYED BIVARIATE SYMBOLS

| (a) | | Uncertainty Representation | | | | | | |
|---|---|---|---|---|---|---|---|---|
| Thematic Representation | Mark | Saturation | Blur | Transparency | Value | Size | Texture | Density |
| | Value | **I** | **S** | **I** | **I** | **S** | ■ | ■ |
| | Size | **S** | **S** | **S** | **S** | ■ | ■ | ■ |
| | Texture | ■ | ■ | ■ | ■ | ■ | ■ | ■ |
| | Density | ■ | ■ | ■ | ■ | ■ | ■ | ■ |
| (b) | | Uncertainty Representation | | | | | | |
| Thematic Representation | Line | Saturation | Blur | Transparency | Value | Size | Texture | Density |
| | Value | I | S | I | I | S | S | A |
| | Size | S | S | S | S | A* | A* | A* |
| | Texture | S | S | S | S | A* | ■ | ■ |
| | Density | S | S | S | A | A* | ■ | ■ |
| (c) | | Uncertainty Representation | | | | | | |
| Thematic Representation | Area | Saturation | Blur | Transparency | Value | Size | Texture | Density |
| | Value | **I** | **S** | **I** | **I** | **S** | S | **A** |
| | Size | **S** | **S** | **S** | **S** | **C** | ■ | ■ |
| | Texture | S | S | S | S | ■ | ■ | ■ |
| | Density | **S** | S | **S** | **A** | ■ | ■ | **C** |

I = Integral, A = Asymmetric, S = Separable, C = Configural. Values underlined and in bold have been established within literature; all others are estimates by this author. *Classification is uncertain and not recommended for employment.

TABLE 4
OPERATIONAL TASKS AND THEIR PERCEPTIVE REQUIREMENTS

| Task | Description | Required Perception |
|---|---|---|
| | Univariate | |
| Identify | Object value lookup | Selective* |
| Compare Within | Compare objects based on value – requires 'Identify' | Ordinal |
| Rank Compare | Compare objects based on value order | Ordinal |
| Ratio Compare | Compare objects based on proportional difference | Quantitative |
| Locate | Isolate object/s based on value | Selective* |
| Distribution | Characterise spatial distribution of all objects equally | Associative |
| Weighted Distribution | Characterise spatial distribution of all objects with weighting based on value | Dissociative |
| | Bivariate | |
| Isolate | Atomistic interpretation of attributes | Separable |
| Compare Between | n-wise attribute comparison for object – requires 'Identify' | Separable |
| Correlate | Determine the relationship between attributes | Integral |
| Associate | Holistic interpretation of attributes | Integral |
| Prioritised Interpretation | Prioritise attribute in holistic interpretation of variables | Asymmetric |
| Weighted Interpretation | Utilise attribute to weight the interpretation of another | Dissociative |
| Associate & Isolate | Holistic and atomistic interpretation of attributes | Configural |
| Combine | Encode attributes into univariate representation | Configural |

*To be selective inherently requires discrete representation of data such that the number of discrete bins sufficiently supports inter-level perceptibility.

the onus falls on the cartographer to determine which operational tasks can and/or should be supported within a given map. This is discussed further in Section 5.1.

If after this process there are still numerous suitable bivariate schemes to choose from, as proposed by step 5 in the design framework, the cartographer should make a final informed decision in adopting a suitable bivariate visualisation. To do so, they may make use of comparative methods and/or other pertinent information (such as dominance principles and measures of semiotic fit discussed in Sections 2.2 and 2.3, respectively). This step is further illustrated in Section 4.3 as part of the case study.

## 4 CASE STUDY

### 4.1 Background

Many adopted farming practices present challenges to decision makers when trying to balance productivity and ecological consequences [1]. For example, farming management decisions impact nutrient and sediment pollution into the Great Barrier Reef (GBR), including nitrogen losses from sugarcane cropping and soil erosion from overland flow [6], [14]. With high variability of rainfall and climate across small areas, and as drought breaking rainfall events produce the highest sediment concentrations, understanding sediment pollution is of particular importance with relation to land use and climate change effects [6], [18].

To date there has been a strong focus on improving spatio-temporal models for use in sediment runoff management, including works by Gladish et al. [14] and Kuhnert et al. [18]. To understand the effects of agricultural management on the GBR, Gladish et al. [14] used a Bayesian hierarchical modelling approach to quantify sediment load and its uncertainty, across various time scales, within the Burdekin catchment. In this context, Kuhnert et al. [18] investigated how uncertainty information represented with exceedance probabilities (linked to thresholds of concern, namely, 837 and 2204 mg/L) can be used for decision-making by highlighting areas where more data are needed to improve load estimates, which areas are hotspots for load delivery, where remediation efforts are working or failing, and which areas are or are not meeting targets.

To showcase the framework, we consider the problems investigated by Kuhnert et al. [18], namely a) flagging hotspots and areas of concern for management intervention and b) assisting with the design and implementation of future monitoring programs, for the Upper Burdekin drainage subdivision. In this paper, instead of exceedance probabilities, these problems are investigated using explicit uncertainty communication methods, specifically through visualisation of suspended sediment concentrations with associated coefficient of variation (see Fig. 3.).

Data utilised was sourced from Gladish et al. [13], and included areal sub-catchment estimates for the 1996 financial year of:
- Total suspended sediment concentration (TSS) in mg/l (posterior mean)
- TSS error in mg/l (posterior standard deviation)

The coefficient of variation (CV) is calculated as the standard deviation divided by the mean.

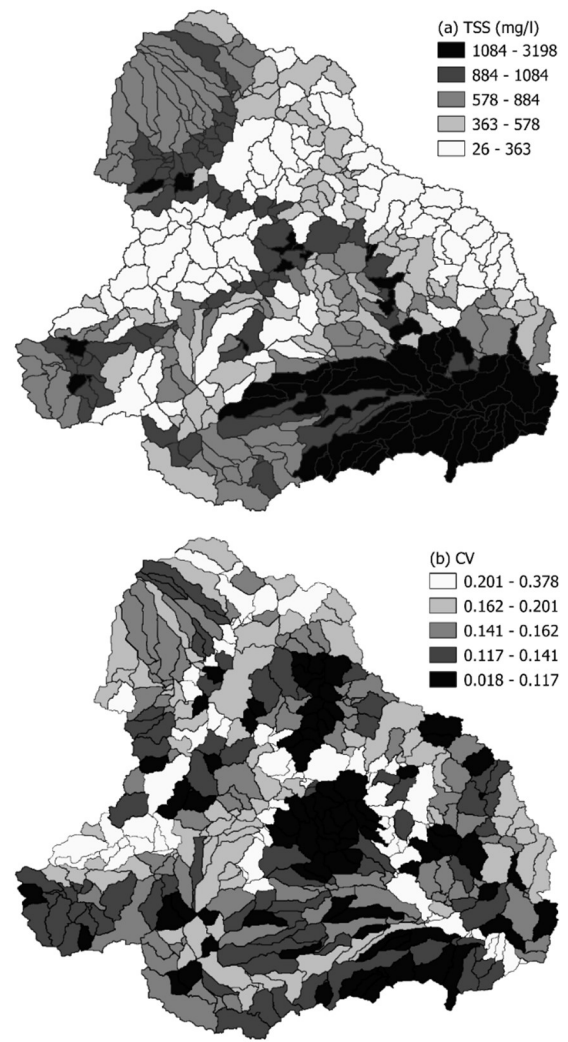

Fig. 3. Case Study Data (a) Total Suspended Sediment (mg/l) and (b) Coefficient of Variation

### 4.2 Problem Definition

When determining areas of concern for management intervention, risk-based decision making considers consequence and likelihood together. Consequently, we need to comprehend the spatial qualities of sediment concentration and related uncertainty across the sub-catchments together, noting that uncertainty may be used to infer likelihood of a consequence. Additionally, for design and implementation of future monitoring programs and remediation activities we are independently concerned both about areas of high concentration, as well as about areas of low confidence in existing estimates.

Therefore, using the framework, an ensemble of comparable visualisations was created, supporting goals of:
a) Drawing attention to high-risk areas (that is, areas of high concentration, where certainty is also high) for ongoing monitoring.
b) Determining areas for implementation of remediation activities based on existing sediment conditions.
c) Prioritising areas for further investigation based on decision confidence (that is, estimate certainty).

Additionally, to highlight how continuousness properties of the representation impacts the utilisation of information, three different binning schemes were included within the ensemble, namely:

1. Classification of TSS into three bins based on thresholds noted in the preceding section, in conjunction using a continuous representation of CV,
2. Five quantile bins of TSS and CV,
3. Classification of TSS into three bins based on thresholds noted in the preceding section, in conjunction with three quantile bins for CV.

Binning schemes 1-3 address the resolution of goals a-c differently. Scheme 1 utilises thresholds from existing literature to justify focus on areas which reach above average or extreme conditions. Scheme 2 facilitates resourcing prioritisation in areas for further investigation and considers the recommendation by Slocum [35] for classifying ordinal data, therefore using quantiles to classify uncertainty. Scheme 3 adopts facets of both.

### 4.3 Framework Application

In employing the framework to address the user goals specified above, the first step was to determine the potential operational map tasks and relate them to their required perceptive properties; initial postulations are presented in Table 5.

By utilising information presented in Tables 2-5, steps 2-4 of the framework design methodology were completed in concert, producing the preliminary set of bivariate recommendations shown in Table 6.

All variables available for thematic representation (value, size, texture and density), are suitable in representing TSS, as they are all ordinal and selective, as well as all having selective lengths of at least five (which make them suitable for addressing all adopted binning schemes listed in Section 4.2). Texture, however, is not dissociative and therefore not applicable for representing CV within this scenario. Of those remaining bivariate combinations, only separable ones are retained.

Step 5 of the framework allows for refinement of the preliminary set of recommendations based on other relevant information. This includes the acknowledgement that uncertainty of the model is not the focus of the research, but rather ancillary to the suspended sediment level estimates. As such, CV should not visually 'dominate' over TSS. This requirement is notable because the need for separability does not preclude visual variables, within a bivariate symbology, from having a meaningful perceptive ordering. As discussed in Section 2.2, dissociative cues visually dominate [2]. Consequently, due to the existing requirement that CV must be dissociative, so too should TSS.

As there are still numerous potential bivariate schemes, further consideration is given to the intuitiveness, performance, and user preference of potential CV representations. Summarising information presented in Section 2.3, for visualising uncertainty, a) blur followed by value then size are the three most intuitive cues, b) value followed equally by blur and transparency have the best performance, but c) saturation followed blur and then value are preferred by users. Due to this incongruity, a selection of these available bivariate symbologies should be directly compared and evaluated. Therefore, out of the remaining applicable bivariate symbologies, the schemes listed in Table 7 were adopted for further consideration.

To finalise a semiotic recommendation, Step 5 of the framework also suggests assessing operational effectiveness of the available bivariate symbologies. Therefore, each scheme in Table 7 was visualised and presented within Fig. 4 for further evaluation. Results are discussed below in Section 4.4.

### 4.4 Results

Fig. 4a appears to be well suited to achieving goal 'a' by drawing attention to areas where both TSS and certainty are high. In contrast, the interpretation of either dimension of information becomes difficult due to the colour mixing, which is occurring simultaneously with both the size and transparency variation. Therefore, this bivariate symbology is seemingly not adequately separable for achieving goals 'b' and 'c'.

When comparing Fig. 4b and 4c (in which pattern size with value is used, but where 4b has value applied directly to the pattern and 4c has pattern overlayed on a value-based choropleth), both retain satisfactory separability. In

TABLE 5
PERCEPTIVE REQUIREMENTS FOR ACHIEVING USER GOALS

| Variable/s | Potential Operational Task | Goal/s Addressed | Required Perception |
|---|---|---|---|
| TSS | Rank Compare | a, b | Ordinal |
| | Locate | a, b | Selective |
| CV | Rank Compare | a, c | Ordinal |
| | Weighted Distribution | c | Dissociative |
| | Weighted Interpretation | a | Dissociative |
| TSS & CV | Isolate | b, c | Separable |

TABLE 6
PRELIMINARY BIVARIATE SYMBOLOGY RECOMMENDATIONS

| | | Uncertainty Representation | | | | | | |
|---|---|---|---|---|---|---|---|---|
| | Area | Saturation | Blur | Transparency | Value | Size | Texture | Density |
| Thematic Representation | Value | | S | | | S | S | |
| | Size | S | S | S | S | | | |
| | Texture | S | S | S | S | | | |
| | Density | S | S | S | | | | |

S = Separable.

TABLE 7
ADOPTED SYMBOLOGY/S ACROSS DIFFERENT BINNING

| | | | Visual Variable (No. Bins) | |
|---|---|---|---|---|
| Binning Scheme | Map Label | Description | TSS | CV |
| 1 | a | Pattern with variable size and transparency | Size (3) | Transparency (Continuous) |
| 2 | b | Pattern with variable value and size | Value (5) | Size (5) |
| | c | Pattern of variable size (black) overlayed on fill of variable value | Value (5) | Size (5) |
| 3 | d | Fill with variable value and blur | Value (3) | Blur (3) |
| | e | Pattern with variable value and size | Value (3) | Size (3) |
| | f | Pattern with variable size and value | Size (3) | Value (3) |

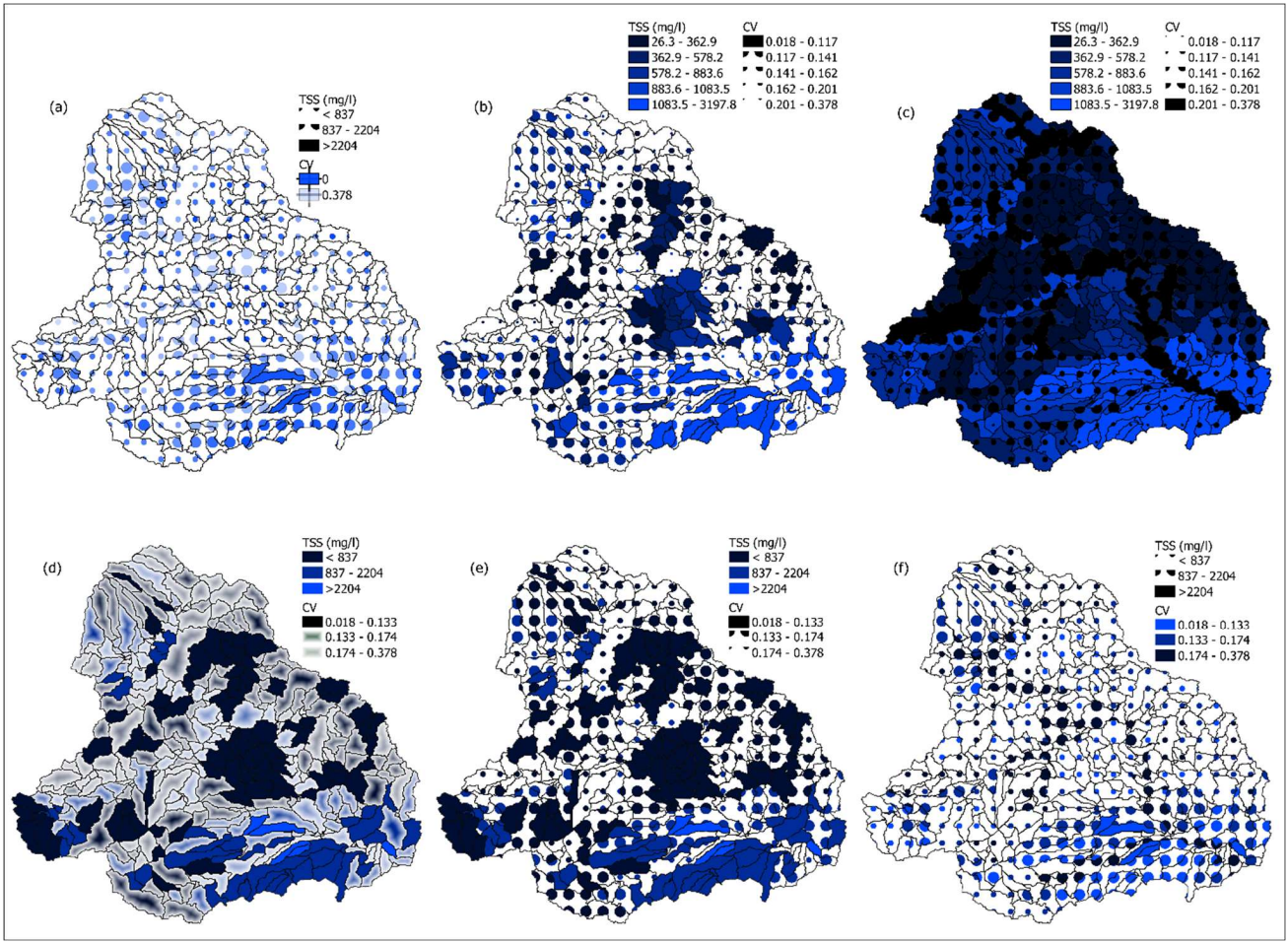

Fig. 4. Map Ensemble Results; (a) Pattern with TSS as Size (3) and CV as Transparency (Continuous), (b) Pattern with TSS as Value (5) and CV as Size (5), (c) Fill with TSS as Value (5) plus Black Pattern Overlay with CV as Size (5), (d) Fill with TSS as Value (3) and CV as Blur (3), (e) Pattern with TSS as Value (3) and CV as Size (3) ,(f) Pattern with TSS as Size (3) and CV as Value (3).

this example, 4c retains greater separability between the two data dimensions, but 4b retains more distinguishability between each bivariate level. This may be due to the difference in colour contrast of red with either white or black. Consequently, 4b is more suited to achieving goal 'a', while 4c is more suited to goals 'b' and 'c'.

As expected, it is notably more difficult to interpret the information from Fig. 4b compared to 4e as there is an increase from 3 levels per dimension in 4e to 5 levels in 4b. This is not just because of the univariate perceptibility limitations, but also the interference/interactions of the visual dimensions, making both the bivariate and univariate levels less discriminable.

Fig. 4d and 4e produce a similar visual distribution of information, but blur is difficult to discern across levels, so is less suitable for identifying values of uncertainty. Conversely, blur is a less distracting and potentially a less interfering cue, more clearly representing the presence or absence of uncertainty.

The application of the size and value conjunction is an interesting case, as shown when comparing Fig. 4e and 4f. In this example, 4e retains a sufficient level of separability; 4f however draws attention to CV over TSS, making 4f an unsatisfactory option.

Overall, of this ensemble, 4a and 4f are unsuitable design options. Of those remaining, the design method demonstrated in 4c is recommended as it a) retains sufficient separability between TSS and CV, b) provides good distinguishability between univariate and bivariate levels, c) is easy to implement in standard GIS software, and d) allows for accomplishment of all potential user goals.

## 5 DISCUSSION

This paper has enhanced decision-making outcomes by improving communication of spatio-temporal data and associated uncertainty. Currently there is no consensus on best practices for bivariate uncertainty visualisation due to variances in user abilities, preferences and requirements [31], [34]. In response to this problem, as suggested by Kinkeldey [15], a semiotic framework has been developed as a mechanism for designing suitable bivariate uncertainty visualisations.

The remainder of this section describes the significance and limitations of the developed framework, including its transferability to other research scenarios and design problems. Also highlighted are notable gaps in knowledge of bivariate semiotics. Finally, results of the case study are related to the broader complexities of bivariate symbol design.

## 5.1 Framework Significance and Limitations

The framework is structured around a five-step design methodology which simultaneously considers the characteristics of the data to be presented and the end-user's goals [17], [25]. As both considerations are requisites in all design problems, this methodology is inherently transferable.

What makes this framework notable is the incidental development of clear associations between potential operational tasks and the consequent required perceptive properties of the visualisation (Table 4). This compendium of information allows for a more effective approach to bivariate map design, while also being transferable to other design problems.

It is impractical to address all possible design considerations within a single framework. Therefore, given our research aims, a set of visual variables and bivariate symbologies, and their corresponding perceptive information, were specially curated to be employed within the proposed methodology. To make use of this framework in other contexts, different sets of employable design options and corresponding background information would likely be required.

To support proper utilisation of the framework, regardless of context, the following considerations are also required.

Firstly, as discussed in Section 3.1, no map is necessarily perfect at achieving all intended user goals. In these instances, it might end up being better to create two maps with a clearer picture for each interpretation requirement than trying to achieve all the goals within one visualisation [38].

Additionally, sometimes the characteristics of the data to be presented doesn't always support the requirements of the visualisation. For example, selective perception might be required to address a specific user goal, but the data itself is classified into too many levels for any viable symbology to retain sufficient distinguishability between levels. The data could be transformed into a lesser number of classes; however, this is not always contextually appropriate and might require a more in-depth investigation into the consequent interpretation of the data. Instead, a complex or redundant cue (such as a multi-hued colour pallet or conjunction of size and spacing, respectively), could be considered to increase the discriminability between levels and variables. Bertin [2] provides an informative dialogue on the perceptive characteristics and use of complex/redundant symbology.

## 5.2 Gaps in Knowledge of Bivariate Semiotics

In developing the framework, some gaps in knowledge of bivariate semiotics were also highlighted.

Notably, it was observed that despite all the research by Nelson [27], Retchless and Brewer [30] and Elmer [10], many bivariate schemes are not yet empirically tested, let alone characterised on the integral-separable spectrum. The current literature on bivariate line symbology is especially limited.

It is possible to infer or derive information from existing characterisations, as was done in completing Tables 2 and 3. Doing so, however, may not always be a straightforward process, particularly as some perceptive properties, namely selective attention, have fuzzy boundaries.

In particular, it is recognised that asymmetric combinations retain some degree of separability [10], and thus could be applicable to scenarios requiring selective attention (such as the case study presented here). Therefore, it could be interesting to investigate where the line between separability and asymmetry is drawn, what degree of separability is retained by various asymmetric representations, and how these factors relate to task performance.

An additional aspect of bivariate symbol design with significant gaps in knowledge is selective length, a property expressly relevant to tasks which involve elucidation of attribute values [36].

It is known that with any more than nine levels of information the distinguishability between steps of a bivariate colour scheme is significantly limited, preventing the scheme from being internalised [19]. Beyond this however, there is minimal understanding of the way different cue interactions affect a bivariate symbol's selectivity, or the selectivity of its constituent dimensions. Likewise, there is minimal understanding of the extent of these effects.

## 5.3 A More In-Depth Look into the Results

Our investigation into the case study confirms the importance of considering user goals in the design stage of creating visualisations. This was highlighted in Fig. 4a and 4f, which were visualisation methods that satisfied the characteristics of the data but were unsatisfactory at supporting the achievement of user goals.

Also observed was the importance of perceptual ordering of cues, regardless of the bivariate symbols standing on the integral-separable continuum. This is highlighted in Fig. 4e and 4f, which have the same bivariate cue combination, value with size, but used in an inverse manner. It was noticed that though both cues were dissociative [2], and the bivariate combination was separable [27], the information represented by value seemed to draw most attention. This contrasts with Bertin's [2] notion that size dominates any other variable combined with it.

Without further empirical testing we cannot confirm whether this is because value in fact has a higher perceptual ordering than size, or if this occurrence is due to another mechanism of the visualisation (such as its spatial distribution). Overall, it does demonstrate a need for further investigation into factors affecting selective attention and perceptual ordering, and their inherent relationship.

Also seen across our resultant ensemble is the occurrence of colour fusion, the qualities of which are directly associated with characteristics of each visual dimension and other maps elements such as the background. Where colour mixing occurs, the discriminability across data dimensions and between levels changes.

Colour fusion is an extremely complex topic, so there is no intent to cover it here, but a few specific occurrences of it in the results can be examined.

In Fig. 4a, there is an unconventional use of transparency. Rather than using it to obscure/expose background data elements, it has been used as an incidental method of colour manipulation. Based on the difficulty of its interpretation within this application, it is suggested that transparency be better suited to convey uncertainty when used in a fog-like manner as described by MacEachren [20].

All maps in Fig. 4 also demonstrate either black or white mixing, the propensity and strength of which change with respect to the size or blur level imposed on the symbol. The dissociative nature of size and blur affect these levels the most, but it also highlights how in this bivariate context the selective length and corresponding choice of extremity representations of a symbol can affect the distinguishability between levels, and across visual dimensions [2].

## 6 CONCLUSION

Visualisations can be used to communicate relationships between modelled spatio-temporal data and associated uncertainty. However, due to differences in user abilities, preferences and requirements, there is no consensus on best presentation methods.

Characteristics of data to be presented help shape the range of available bivariate symbols, but ultimately, the utility of a map depends on the achievement of the end-user's goals. Therefore, the characteristics of the data, and goals and abilities of the user, should be considered in concert when designing a visualisation.

These issues were addressed in this paper by development of a data- and task- oriented visualisation framework which can be used in the bivariate communication of uncertainty. In doing this, a transferable design methodology and set of associations between potential operational tasks and perceptive requirements of maps were incidentally created.

Some gaps in knowledge of bivariate symbol design were also observed, highlighting the need for further investigate into certain aspects of visual variable interaction. Specifically, these included the need for better understanding of the discernability of information under different visual conditions and defining the relationship between perceptual ordering and selective attention.

## ACKNOWLEDGMENT

The authors wish to thank Petra Kuhnert from CSIRO's Data61 for providing the motivating example for this research.